\documentstyle[11pt,paspconf]{article}

\begin{document}

\title{Probing the Galactic Halo with Globular Cluster Tidal Tails}

\author{Carl J. Grillmair} \affil{Jet Propulsion Laboratory, Mail Stop
183-900, 4800 Oak Grove Drive, Pasadena CA 91109}

\begin{abstract}
We discuss recent observational and numerical work on tidal tails in
globular clusters. Evidence of tidal tails has now been found in 16
Galactic globulars and 4 globulars in M31. Simulations indicate that
mapping of these tidal tails over most of a typical cluster's orbit
may be quite feasible. A relatively modest effort in this regard would
yield reliable space velocities for those clusters whose tidal tails
could be traced to angular separations of a few degrees or more.
Mining of forthcoming databases should allow us to trace the
individual orbital paths of globular clusters over large portions of
the sky. This would enable us to better constrain models of the
collapse of the Galaxy and to build up an accurate picture of the
distribution of matter in the Galactic halo.
\end{abstract}



\section{Introduction}

Globular clusters have long been known to be limited in extent by the
tidal stresses imposed by the potential field of the Galaxy. King
(1966) developed a remarkably successful model for globular clusters
in which the spatial extent of a cluster (the ``tidal radius'' =
$r_t$) is determined by the strength of Galactic tidal forces at the
cluster's perigalacticon. Extensive campaigns to measure the radial
surface density profiles of globular clusters using aperture
photometry and star counts (e.g. Kron \& Mayall 1960; King et
al. 1968; Peterson 1976, Hanes \& Brodie 1985) determined that the
form of the profiles appeared to be in agreement with King models over
several orders of magnitude for almost all clusters observed.

The direct relationship between the tidal radius and the orbital
periGalactic distance assumed by King prompted other investigators to
try to measure individual globular cluster orbit shapes (or conversely
the potential field of the Galaxy) using the tidal radii computed from
fits of King models to the data (e.g. Peterson 1974; Innanen, Harris,
\& Webbink 1983). However, it was soon realized that the uncertainties
in both the published profiles and in our understanding of weak tidal
encounters severely hindered this type of analysis.  With the advent
of parallel computer architectures, more efficient N-body codes, fast
plate-scanning machines, and large-format CCD detectors, great strides
have been made on both these fronts. It may not be altogether
surprising that many of the questions we had once asked may turn out
to be largely irrelevant to the problems we ultimately want to solve.

\section{Numerical Work}

Globular cluster evolution and the process of tidal stripping have
been addressed many times both analytically and numerically (see for
example Spitzer 1987 and references therein; Lee \& Ostriker 1987;
Allen \& Richstone 1988, McGlynn 1990; McGlynn \& Borne 1991). Many
works have focussed on the stability of various types of stellar orbits
under the influence of tidal forces, and the consensus is that the
short-term effects of tidal processes on clusters are considerably
more complicated that the simple binding energy cutoff assumed in the
King model. The question of whether $r_t$ accurately predicts the
eventual limiting radii of clusters is still a source of some
contention, with proponents favoring values ranging from $0.5r_t$ to
$r_t$.

\begin{figure}
\vspace{3.0in}
\includegraphics{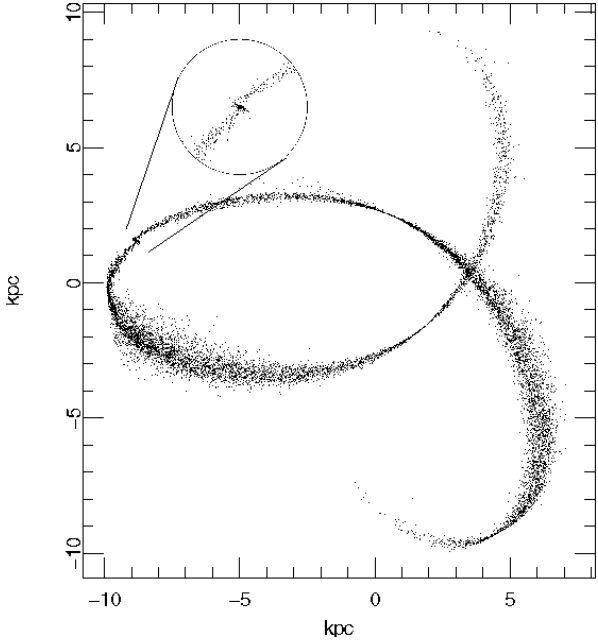}
\caption{A 64K-particle cluster in a spherical, logarithmic
potential (centered at the origin) after 30 orbits. The cluster (at
left center) survives despite having lost nearly 75\% of its initial
mass. The inset shows the region near the cluster magnified to reveal
the signature plumes of the most recent perigalactic passages.}
\label{fig:model} 
\end{figure}

More recent work on weak tidal encounters utilizing both the
Fokker-Planck approach (Oh \& Lin 1992; Lee \& Goodman 1995) and
self-consistent, large-scale N-body techniques (Grillmair et al. 1998)
confirm that tidal stripping is not a very efficient process and that
only a fraction of unbound or marginally-bound particles are removed
in any particular tidal encounter. Combined with continual two-body
relaxation in the core of the cluster and the consequent replenishment
of the region near $r_t$, a long-term, episodic flow of particles away
from the cluster should be expected.  Given the very low escape
velocities near $r_t$, stripped particles may remain in the vicinity
of the cluster for several galactic orbits. Depending on which end of
the equipotential surface the escaping particles passed through, they
would then either migrate ahead of the cluster or fall behind, giving
rise to slowly growing tidal tails.  Figure \ref{fig:model} shows one
of Grillmair et al.'s model clusters after 30 eccentric orbits in a
spherical, logarithmic potential.

The effects of a disk were not considered in the works cited above,
but disk shocking of the cluster is expected to variously enrich,
broaden, and lengthen the tidal tails as a result of heating of the
outer parts of the cluster (e.g. Murali \& Weinberg 1997).  In
addition, irregularities in the Galactic disk such as giant molecular
clouds will tend to scatter stars already in the tidal
tails. Nonspherical galactic potentials would similarly tend to
broaden the tidal tails as the orbits of stars precess independently
of one another. Further modeling work should be aimed at studying the
observable consequences of these processes, of using realistic stellar
mass distributions (e.g. Johnstone 1993; Lee \& Goodman 1995), and at
better addressing the survival statistics of globular clusters in the
face of continual weak tidal encounters.

\section{Observations}

\begin{figure}[t]
\vspace{2.8in}
\includegraphics{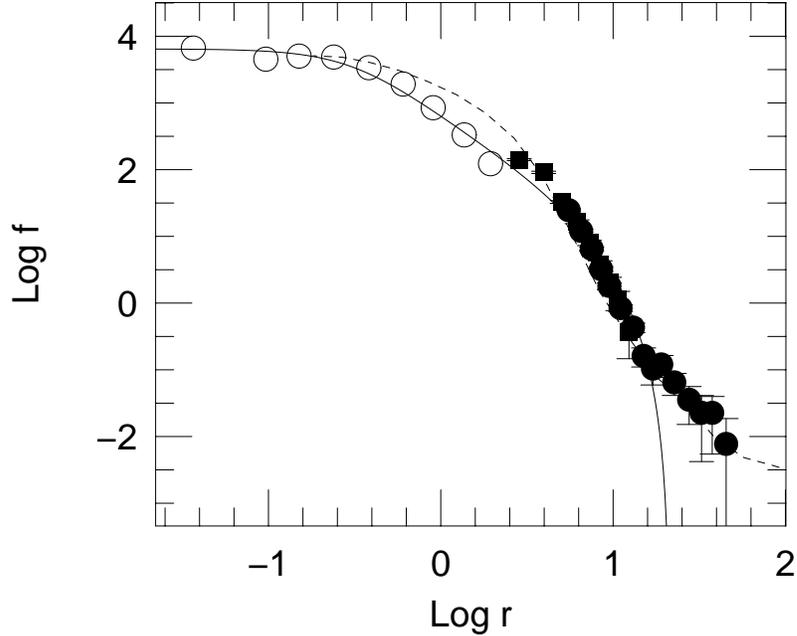}
\caption{Surface density profile of NGC 7089 (=M2). Open circles
represent aperture photometry of Hanes \& Brodie (1985), filled
squares and filled circles show the star counts of King et al. (1968),
and of Grillmair et al. (1995), The best-fit King model is shown by
the solid line, and the dashed line shows one of Grillmair et al.'s
(1998) models arbitrarily normalized to the data at large
$r$. Departures at small radii reflect the differences between King
models and the Jaffe model used in the numerical experiments. Note the
break near the tidal radius and the power-law decline at large $r$
characteristic of tidal tails.}
\label{fig:n7089} 
\end{figure}

\begin{figure}
\vspace{3.0in}
\includegraphics{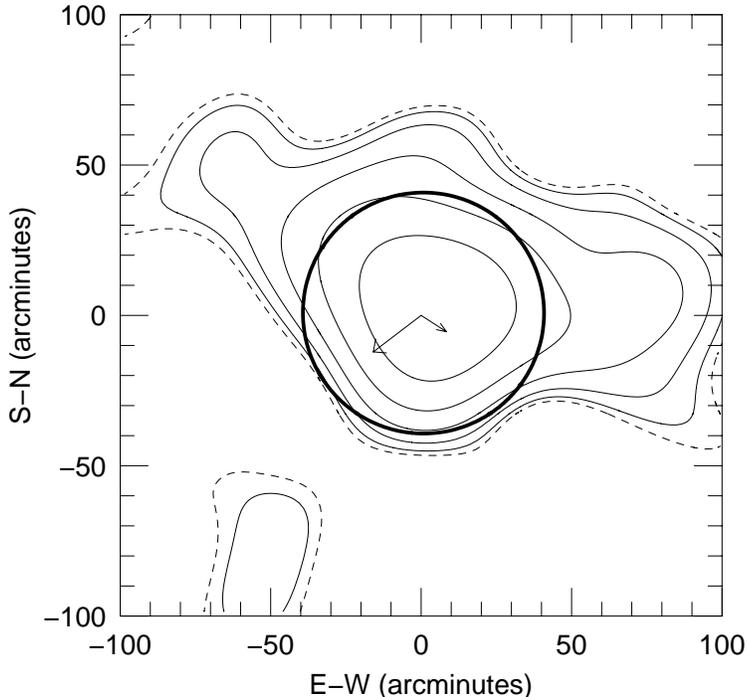}
\caption{Contours of the color-magnitude-selected star
distribution around NGC 7089 after smoothing with a Gaussian kernel
with $\sigma = 20$ arcminutes. The heavy line shows the extent of the
best-fit King model after the same degree of smoothing and at a
surface density level equal to that shown by the dashed
line.}
\label{fig:2d} 
\end{figure}

Early work on defining the limits of globular clusters through star
counts were limited by the $\sqrt N$ uncertainties introduced by the
large number of foreground and background stars. Using photographic
plates taken in two colors, Grillmair et al. (1995) were able to
reduce this foreground/background contamination by 90\% or more by
selecting stars based on their colors and magnitudes. Most of the
globular clusters in their sample were found to depart from the form
predicted by King models, with an excess surface density of stars at
large $r$ and significant numbers of stars with $r > r_t$. Figures
\ref{fig:n7089} and \ref{fig:2d} show the one and two-dimensional
distributions of color-selected stars around NGC 7089. The departures
in the surface density profiles are similar both qualitatively and
quantitatively to the numerical results, with a ``break'' from a King
profile at $r < r_t$ followed by a powerlaw decline which varies from
cluster to cluster. This variation in the power-law component is
consistent with the modeling results of Grillmair et al. (1998), who
found that the gradient in the extended surface density profile
depends both on the shape of the cluster's orbit and, through
conservation of energy and angular momentum, on the orbital phase of
the cluster. At some level (which depends on the rate of evaporation of
stars and on the orbit's orientation with respect to our line of
sight) the tidal tails will obscure the tidal cutoffs predicted by
King-Michie models.

Evidence of tidal tails has now been found in 16 Galactic globular
clusters, including M2, M15, NGC 288, NGC 362, NGC 1904, NGC 4590, NGC
5824, NGC 6864, NGC 6934, NGC 6981 (Grillmair et al. 1995), M5, M12,
M13, M15, NGC 5466 (Lehmann \& Scholz 1996), M5 (Kharchenko, Scholz,
\& Lehmann 1997), M55 (Zaggia, Piotto, \& Capaccioli 1997), and
$\omega$ Cen (Leon \& Meylan 1997). Efforts are underway to extend
this sample, and to trace known tidal tails beyond the $2\deg$ or less
to which existing data have been limited.  Interestingly, tidal tails
have recently also been detected in four globular clusters in M31
(Grillmair et al.  1996; Holland, Fahlman, \& Richer 1997). The
decline in the number of blue foreground stars at $V > 23$, combined
with both a large projected field of view and spatial resolution
sufficient to distinguish between stars and galaxies, makes
this a fairly straightforward endeavor using the Wide Field Planetary
Camera 2 on the Hubble Space Telescope.

Given the faintness of the stars which make up the bulk of the tidal
tails and the fact that they constitute only a small fraction of the
color-selected field stars, obtaining spectra would be more than a
minor undertaking. Nonetheless, spectra would be useful both to
confirm association of these stars with their parent clusters and, by
examining their velocity distributions, to study the interaction of
the cluster and Galactic potentials. Indeed, in a sample of 237 stars
in the outskirts of M15, Drukier et al. (1998) find that the velocity
dispersion appears to have a minimum at about 7 arcminutes, well
inside the 23 arcminute tidal radius of the cluster.  These authors
favor tidal heating as being responsible for a subsequent slight rise
in the velocity dispersion. It is also possible that their sample
includes a small number of stars residing in the unbound halo or the
tidal tail and projected along our line of sight.

Interestingly, the tidal tails observed to date imply that globular
clusters cannot be surrounded by halos of dark matter. Moore (1996)
showed for NGC 7089 that the existence of the tidal tails requires a
global mass-to-light ratio of $M/L < 2.5$, consistent with values for
$M/L$ derived for the central regions of globular clusters.

\section{Tracing Cluster Orbits}

\begin{figure}
\vspace{2in}
\includegraphics{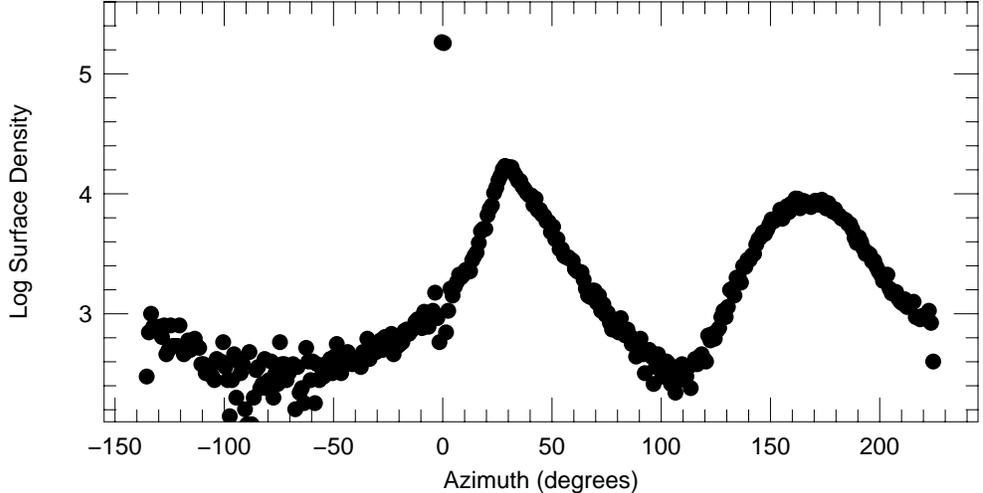}
\caption{The number of stars per square degree down to $V = 23$ measured
along the tidal tails of a simulated cluster. The cluster is viewed at
some arbitrary point along an orbit inclined $45\deg$ with respect to
the plane of the Galaxy, and the surface densities are scaled to match
the number of {\it cluster} stars in NGC 7089.}
\label{fig:tail}
\end{figure}

The presence of tidal tails give us a potentially useful tool for
determining cluster orbits and mapping the Galactic potential
field. Mining large data sets like the Guide Star Catalog II will soon
make it possible to trace tidal tails to large angular distances from
their parent clusters. Figure \ref{fig:tail} shows the surface density
distribution in one of Grillmair et al.'s (1998) 64K-particle models
as it might appear on the sky from our Galactic vantage point. The
surface density of particles has been scaled to approximately match
the observed number of stars in NGC 7089. If, as is the case for this
particular model, NGC 7089 has lost the majority of its original stars
to its tidal tails, we would expect to find $\sim 10^6$ stars brighter
than $V = 23$ strewn along this single cluster's orbit.  The highest
surface densities in Figure \ref{fig:tail} (with the exception of the
cluster itself) correspond to an apogalactic portion of the cluster
orbit and to projection of the tail along our line of sight. With
suitably optimized color-selection criteria, and with the exception of
regions very near the Galactic plane, it would be quite feasible to
follow such tails for most of their length. 

More challenging will be the issue of confusion. There are probably
many hundreds of apogalactic, high-density swarms of blue turnoff
stars laid down by old halo clusters. Determining their
orientations in the presence of contamination by unaffiliated halo
stars and unresolved galaxies and matching such tail segments with
parent clusters may well not be trivial.  Simulations show that the
solution will most often be to obtain the deepest photometry possible,
both to minimize photometric errors and thereby keep the
color-magnitude selection criteria as restrictive as possible, and to
take advantage of the fainter end of the globular cluster luminosity
function.

\begin{figure}
\vspace{2.5in}
\includegraphics{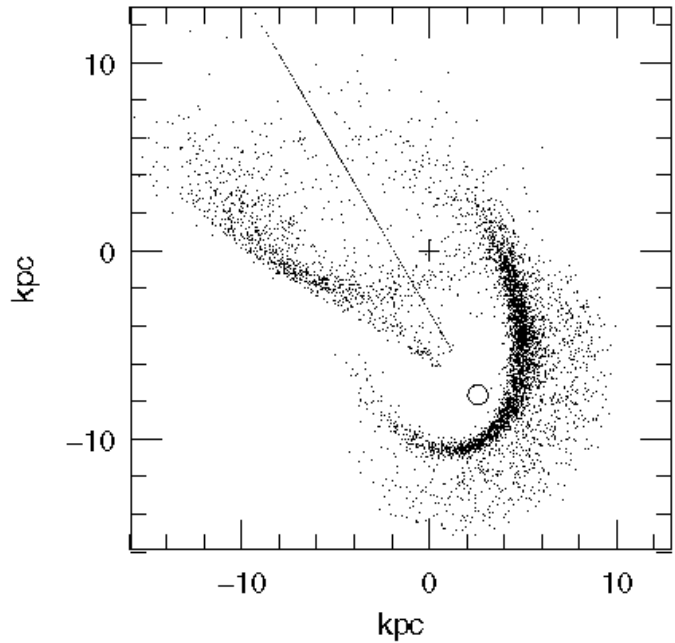}
\caption{Simulation of a typical orbital trace using photometric
distances to tidal tail stars. The open circle denotes the location
(projected onto the plane of the cluster's orbit) of our solar system,
and the Galactic center resides at the origin. The cluster's position
is indicated by the ``finger of God'' resulting from photometric
uncertainties. Note that the effects of sample contamination by field
stars (which will vary hugely from one orbit to the next) are not
included here.}
\label{fig:recovered}
\end{figure}

Another reason for going deep is to reduce the uncertainties in
photometrically determined distances.  While the turnoff regions of
most globular clusters stand out rather well from the distributions of
Galactic field stars, this region is less useful for determining
relative distances via main sequence fitting. Figure
\ref{fig:recovered} shows a simulation based on a model of Grillmair
et al. (1998) wherein particles have been assigned colors and
magnitudes using a typical metal-poor globular cluster isochrone and
luminosity function. Assuming reasonable exposure times on a 1-meter
Schmidt-type telescope and applying appropriate uncertainties, we
recover photometric distances down to a limiting apparent magnitude of
$V \sim 23$ (or about 3 magnitudes below the turnoff). Despite the
photometric errors and the uncertainty in absolute stellar
luminosities in the region near the turnoff, Figure
\ref{fig:recovered} shows that the cluster's orbit can be readily
traced over most of the extent of the tidal tails.  Globular cluster
orbit shapes can thus be determined on a case by case basis, rather
than with the statistical arguments used in the past.

Knowing a cluster's line-of-sight velocity, the orbital path of the
cluster as deduced from it's tidal tails immediately yields a
reasonably accurate value for the cluster's space velocity; the
orientation of the tidal tails on the sky near the cluster gives the
second component of the velocity and the photometric distances of
stars in the tails yield the third. Conservation of orbital angular
momentum then allows us to compute the space velocity at any point
along the tidal tails (we neglect here the small error introduced by
by the fact that the orbits of stars in the tidal tails do not exactly
coincide with that of the cluster). The halo potential itself must be
determined numerically (e.g. Binney and Tremaine 1987). Using normal
points derived from Figure \ref{fig:recovered} we recover the
spherical model potential to within 10\% over the radial range sampled
by the orbit. Since the Galactic halo potential is non-Keplerian and
is quite probably nonspherical as well, even a well-determined
globular cluster orbit will not be sufficient to map the potential at
this level of accuracy.  However, tracing the orbits of several clusters
would provide strong constraints on the distribution of matter in the
inner portion of the Galactic halo.  Combining these orbit shapes and
mass distributions with similar information gleaned from the tidal
tails of the Milky Way's satellite galaxies (Johnston, Hernquist, \&
Bolte 1996) would greatly improve our understanding of the
initial collapse and subsequent evolution the Galaxy.


\begin{references}

\reference Allen, A. J., \& Richstone, D. O. 1988, \apj, 325, 583
\reference Binney, J., \& Tremaine, S. 1987, {\it Galactic Dynamics},
Princeton University Press: Princeton.  
\reference Drukier, G. A., Slavin, S. D., Cohn, H. N., Lugger, P. M.,
\& Berrington, R. C. 1998, \aj, in press.
\reference Grillmair, C. J., Freeman, K. C, Irwin, M., \& Quinn,
P. J. 1995, \aj, 109, 2553
\reference Grillmair, C. J., Ajhar, E., Faber, S. M., Baum, W. A.,
Lauer, T. R., Lynds, C. R., \& O'Neil, E. Jr. 1996, \aj, 111, 2293 
\reference Grillmair, C. J., Quinn, Freeman, K. C., Salmon, J.,
Prince, T., \& Messina, P. 1998, in preparation.
\reference Hanes, D. A., \& Brodie, J. P. 1985, \mnras, 214, 491
\reference Holland, S., Fahlman, G. G., \& Richer, H. B. 1997, \aj,
114, 1488
\reference Innanen, K. A., Harris, W. E., \& Webbink, R. F. 1983, \aj,
88, 338
\reference Johnston, K. V., Hernquist, L., \& Bolte, M, 1996, \apj,
465, 278
\reference Johnstone, D. 1993, \aj, 105, 155
\reference Kharchenko, N., Scholz, R.-D., \& Lehmann, I. 1997, \aaps,
121, 439  
\reference King, I. R.  1966, \aj, 71, 276
\reference King, I. R., Hedemann, E., Hodge, S. M., \& White, R. E.
    1968, \aj, 73, 456
\reference Kron, G. E., \& Mayall, N. U. 1960, \aj, 65, 58
\reference Lee, H. M., \& Ostriker, J. P. 1987, \apj, 322, 123
\reference Lee, H. M., \& Goodman, J. 1995, \apj, 443, 109
\reference Leon, S., \& Meylan, G. 1997, personal communication.
\reference McGlynn, T. A. 1990, \apj, 281, 184
\reference McGlynn, T. A., \& Borne, K. D. 1991, \apj, 372, 31
\reference Moore, B. 1996, \apj, 461, L13
\reference Murali, C., \& Weinberg, M. D. 1997, \mnras, 291, 717
\reference Oh, K. S., \& Lin, D. N. C. 1992, \apj, 386, 519
\reference Peterson, C. J. 1974, \apj, 190, L17
\reference Peterson, C. J.  1976, \aj, 81, 617
\reference Spitzer, L.  1987, {\it Dynamical Evolution of Globular
Clusters}, Princeton University Press: Princeton. 
\reference Zaggia, S., Piotto, G., \& Capaccioli, M. 1997, \aap, 327,
1004 
\end{references}
\end{document}